\documentclass[fleqn,usenatbib]{mnras}

\usepackage{newtxtext,newtxmath}
\usepackage{mathrsfs}
\usepackage[T1]{fontenc}
\usepackage{ae,aecompl}

\usepackage{graphicx}	% Including figure files
\usepackage{amsmath}	% Advanced maths commands
\usepackage{amssymb}	% Extra maths symbols

\title[SFRs of two GRB hosts at $z\sim 2$ and a C~{\sc ii} deficit observed with ALMA]{Star-formation rates of two GRB host galaxies at $z\sim 2$ and a [C~{\sc ii}] deficit observed with ALMA}

\author[T. Hashimoto et al. 2019]{
Tetsuya Hashimoto,$^{1}$\thanks{E-mail: tetsuya@phys.nthu.edu.tw}
Bunyo Hatsukade,$^{2}$
Tomotsugu Goto,$^{1}$
Seong Jin Kim,$^{1}$
\newauthor
Kouji Ohta,$^{3}$
Tohru Nagao,$^{4}$
Albert K. H. Kong,$^{1}$
Kouichiro Nakanishi,$^{5}$
\newauthor
and Jirong Mao$^{6,7}$
\\
$^{1}$Institute of Astronomy, National Tsing Hua University, 101, Section 2. Kuang-Fu Road, Hsinchu, 30013, Taiwan (R.O.C.)\\
$^{2}$Institute of Astronomy, Graduate School of Science, The University of Tokyo, 2-21-1 Osawa, Mitaka, Tokyo 181-0015, Japan\\
$^{3}$Department of Astronomy, Kyoto University, Kitashirakawa-Oiwake-cho, Sakyo-ku, Kyoto, Kyoto 606-8502, Japan\\
$^{4}$Research Center for Space and Cosmic Evolution, Ehime University, Matsuyama, Ehime 790-8577, Japan\\
$^{5}$National Astronomical Observatory of Japan, 2-21-1 Osawa, Mitaka, Tokyo 181-8588, Japan;\\ 
SOKENDAI (The Graduate University for Advanced Studies), 2-21-1 Osawa, Mitaka, Tokyo 181-8588, Japan\\
$^{6}$Yunnan Observatories, Chinese Academy of Sciences, 650011 Kunming, Yunnan Province, China\\
$^{7}$Key Laboratory for the Structure and Evolution of Celestial Objects, Chinese Academy of Sciences, 650011 Kunming, China\\
}
\date{Accepted 2019 July 19. Received July 17; in original form July 03}

\pubyear{2019}

\begin{document}
\label{firstpage}
\pagerange{\pageref{firstpage}--\pageref{lastpage}}
\maketitle

\begin{abstract}
Event rate of long Gamma-Ray Bursts (GRBs) is expected to be an useful tracer of the cosmic star-formation history.
For this purpose, it is necessary to understand what kind of star formations/galaxies are traced by GRBs.
%Some GRB host galaxies indicate extremely high star-formation rates (SFRs) implied from radio observations.
%However observed radio fluxes could be contaminated by long-lived afterglows of GRBs even several years after the bursts.
%This prevents us from correctly understanding galaxy populations hosting GRBs especially at $z\sim 1-2$.
Here we report rest-frame far-infrared (FIR) continuum detections of GRB 070521 and 080207 host galaxies at $z\sim2$ with ALMA band 8 and 9.
The FIR photometries provide the reliable star-formation rates (SFRs), because FIR emission is free from dust extinction and possible radio contamination from long-lived afterglows of GRBs.
The spectral energy distribution fittings indicate 49.85$^{+72.33}_{-2.86}$ and 123.4$^{+25.19}_{-21.78}$ $M_{\odot}$ yr$^{-1}$ for the 070521 and 080207 hosts, respectively.
The derived SFRs place them on the \lq \lq main sequence\rq \rq\ of normal star-forming galaxies at $z\sim2$.
The derived SFRs are significantly lower than that of radio observations.
It is inferred that the observed radio fluxes in a previous study are contaminated by the afterglows.
ALMA marginally detected [C~{\sc ii}]\,158\,$\mu$m emission line from the GRB 080207 host galaxy with S/N $\sim$ 4.
This is the first detection of [C~{\sc ii}]\,158\,$\mu$m of a GRB host at $z>2$, and the second detection among known GRBs.
The luminosity ratio of [C~{\sc~ii}]\,158$\mu$m to FIR is 7.5$\times 10^{-4}$, which is one of the smallest values among galaxies at $z\sim 1-2$ with the same FIR luminosity. 
The \lq \lq [C~{\sc ii}] deficit\rq \rq\ could be a new physical property to characterise GRB hosts at $z\sim1-2$.
Possible parameters controlling the deficit include the metallicity, initial mass function, and gas density.
\end{abstract}


%%%%%%%%%%%%%%%%%%%%%%%%%%%%%%%%%%%%%%%%%%%%%%%%%%

%%%%%%%%%%%%%%%%% BODY OF PAPER %%%%%%%%%%%%%%%%%%
\section{Introduction}
\label{introduction}
Long Gamma-Ray Bursts (hereafter GRBs) are associated with explosions of massive stars at cosmological distances.
The bright gamma-ray and afterglow emissions allow us to detect GRBs even at the very high redshift, $z\sim$ 8 \citep{Tanvir2009,Salvaterra2009}.
The gamma-ray also can penetrate dust.
Therefore the event rate of GRBs is expected to be an useful tool to trace the cosmic star-formation history \citep[e.g.,][]{Yonetoku2004,Trenti2012}.
For this purpose, it is necessary to understand what kind of galaxies and star formation are traced by GRBs.

At the local Universe, GRB are hosted by galaxies with special characteristics, e.g., faint, blue star-forming, and low metallicity (or low iron-abundance) galaxies \citep[e.g.,][]{Fruchter2006,Savaglio2009,Levesque2010,Hashimoto2018}.
These properties are likely linked to physical conditions required for massive stars to launch relativistic jets of GRBs.
Theoretically GRB progenitor needs a sufficient angular momentum to form an accretion disk and the jet.
A low metallicity environment is favoured because the progenitor does not lose its angular momentum by the mass loss during its evolution \citep[e.g.,][]{Woosley2006}.

At $z\sim1-2$, hereafter \lq \lq high-$z$\rq\rq, an averaged metallicity of star-forming galaxies is lower than that of local galaxies \citep{Savaglio2009,Hayashi2009,Montero2009,Yabe2012,Zahid2014}.
Therefore the theoretical low metallicity requirement for high-$z$ GRBs might be relatively less important.
\citet{Perley2016b} estimated stellar masses of GRB host galaxies at $z\sim1-2$ as an indicator of the metallicity.
They argued that a threshold metallicity for high-$z$ star-forming galaxies to host GRBs is approximately below the Solar value, while the GRB rate in super solar metallicity environments is heavily suppressed.
If the GRB rate is controlled by metallicity and a majority of star-forming galaxies at $z\sim1-2$ show sub-solar metallicity, high-$z$ GRBs could be hosted by more representative star-forming galaxies compared to $z \sim 0$.
GRBs are found in massive red star-forming galaxies at $z\sim1-2$  \citep[e.g.,][]{Hashimoto2010,Hashimoto2015,Perley2016a,Perley2016b}.
This galaxy population is obviously different from the local ones.
This is probably reflecting the cosmic star-formation history in which a fractional contribution of dust-obscured star formation is larger at higher redshift \citep[e.g.,][]{Takeuchi2005,Goto2010}. 
Actually some GRBs at $z\sim1-3$ were localised at ultraluminous infrared galaxies (ULIRGs) that indicate intense dust-obscured star formation \citep[e.g.,][]{Perley2017b}.
Such host galaxies indicate extremely high obscured star-formation rate (SFR) up to $\sim 10^{3}$ $M_{\odot}$ yr$^{-1}$ measured by radio wavelength \citep{Perley2017b}.
High-$z$ GRBs might be able to occur even in such extreme galaxies in contrast to local GRB hosts.
This may be due to the fact that at low-$z$ the fraction of star formation density hosted in ULRIGs is very small \citep[e.g.,][]{Goto2010}.
However, radio emission of GRB host galaxies could be contaminated by long-lived afterglows even several years after the burst.
To avoid the afterglow contamination securely, radio observations need a long time delay of $\sim$ 10 years \citep{Perley2017b}.
Therefore the current SFR estimates by radio observations can be much overestimated, which prevents us from correctly understanding galaxy populations of GRB hosts.

Another problem is a difficulty of host characterisation by UV/optical data such as metallicity measurement.
The metallicity is an essentially important parameter to characterise GRB hosts, because under most theoretical GRB models, formation of a GRB progenitor able to launch a relativistic jet requires a low metallicity environment as mentioned above. 
However the metallicity is measured from rest-frame optical emission lines or UV absorption lines in general. 
Because of the dust extinction, physical parameters derived from UV/optical data are only applicable to non-dusty GRB hosts. 
The number fraction of the dusty GRB host galaxies begins to increase at $z\sim1-2$ \citep[e.g.,][]{Perley2013a}.
Therefore a new \lq \lq dust extinction-free\rq \rq\ parameter to characterise GRB host galaxies at $z\sim1-2$ has been awaited.

[C~{\sc ii}]\,158$\mu$m is one of the brightest of emission lines from far-infrared (FIR) through meter wavelengths emitted by star-forming galaxies, almost unaffected by dust extinction.
IR-luminous galaxies indicate low [C~{\sc ii}]\,158$\mu$m/FIR luminosity ratios, known as \lq \lq [C~{\sc ii}] deficit\rq \rq\ \citep[e.g.,][]{Malhotra1997}.
The [C~{\sc ii}]\,158$\mu$m luminosity is probably controlled by physical conditions of inter-stellar medium including FUV radiation field, gas density, initial mass function (IMF), and metallicity \citep[e.g.,][]{Wolfire1989,Stacey1991,Kaufman1999,Hailey-Dunsheath2010,Stacey2010,Lagache2018}.
Therefore the [C~{\sc ii}]\,158$\mu$m/FIR luminosity ratio could be an useful parameter to characterise physical environments of GRBs at $z\sim1-2$.

In this paper, we report rest-frame FIR detections of (U)LIRGs hosting GRB 070521 and 080207 at $z\sim2$.
Atacama Large Millimeter/submillimeter Array (ALMA) bands covered redshifted FIR peaks of spectral energy distributions (SEDs).
By including FIR, i.e., peak wavelength of dust emission heated by obscured star formation, SED fitting analysis provides reliable obscured SFRs.
The derived SFRs are free from the possible afterglow contamination, because the FIR afterglow contributions extrapolated from observed radio fluxes are negligible compared with the observed FIR fluxes, assuming standard spectral slopes of GRB afterglows \citep{Sari1998} (see Section \ref{comparison} for details).
%The SFRs of 070521 and 080207 hosts are 49.85$^{+72.33}_{-2.86}$ and 123.4$^{+25.19}_{-21.78}$ $M_{\odot}$ yr$^{-1}$, respectively, placing them on the \lq \lq main sequence\rq \rq\ galaxies at the same redshift. 
%ALMA marginally detected [C~{\sc ii}]\,158\,$\mu$m emission of GRB 080207 host with S/N$\sim$4. 
%This is the first detection of [C~{\sc ii}]\,158\,$\mu$m emission of GRB hosts at $z>2$ and the second detection among GRBs known.
%We demonstrate the [C~{\sc ii}]\,158\,$\mu$m deficit of the GRB 080207 host.
%In terms of the deficit as a function of FIR luminosity, the host shares the same location as those in local star-forming galaxies.
%The deficit is highlighted when the host is compared with star-forming galaxies at the same redshift. 

The structure of this paper is as follows.
We describe the sample selection in Section \ref{sample}.
Section \ref{observation} consists of configurations of ALMA observations and data analysis together with observed quantities.
SED fitting analysis based on multi-wavelength data is described in Section \ref{analysis_results}.
We demonstrate physical properties of our sample derived from the SED fitting analysis also in Section \ref{analysis_results}.
SFRs and [C~{\sc ii}]\,158\,$\mu$m deficit are discussed in Section \ref{discussion} followed by conclusions in Section \ref{conclusion}.
Throughout this paper, we assume a cosmology of ($\Omega_{m}$,$\Omega_{\Lambda}$,$\Omega_{b}$,$h$)=(0.307, 0.693, 0.0486, 0.677) by $Planck15$ \citep{Planck2015}, unless otherwise mentioned.
Referred SFRs and stellar masses are based on \citet{Chabrier2003} IMF except for SFRs derived from radio observations \citep{Perley2013b}.
While \citet{Kroupa2001} IMF is adopted in \citet{Perley2013b}, the difference between stellar masses derived from \citet{Chabrier2003} and \citet{Kroupa2001} IMFs is $\sim$8\% \citep{Madau2014}.
This is negligible compared with the uncertainties of the radio observations for our sample \citep[S/N$\lesssim7$;][]{Perley2013b}.

\section{Sample}
\label{sample}
We collected {\it Swift} GRBs localised to ultraluminous host galaxies with spectroscopic redshifts from the literature \citep{Perley2017b}.
Here the ultraluminous host galaxy is defined by either SFR $>$ 100 $M_{\odot}$ yr$^{-1}$ or $L_{\rm IR} > 10^{12} L_{\odot}$.
ALMA has a capability to observe GRB 061121 \citep[$z$ = 1.314;][]{Fynbo2009}, 070521 \citep[$z$ = 2.0865;][]{Kruhler2015}, and 080207 \citep[$z$ = 2.0858;][]{Kruhler2012} host galaxies in rest-frame FIR wavelength range. 
The SFRs derived from submillimeter/radio of these targets are one order of magnitude higher than those calculated from optical wavelength, suggesting strong dust-obscured star formations or possible contamination from radio afterglows. 

ALMA cycle 5 observations (Project code: 2017.1.00337.S) were carried out for the GRB 070521 and 080207 hosts to detect the FIR emission.
The GRB 061121 host was not observed with ALMA, since the cycle 5 semester ended before the completion of the full observations.
We briefly summarise previous studies on individual host galaxy as follows.

\subsection{GRB 070521 host}
GRB 070521 was detected by the {\it Swift} Burst Alert Telescope (BAT) \citep{Gehrels2004,Guidorzi2007a}.
No optical afterglow was detected \citep{Marshall2007,Greiner2007}, suggesting an optically \lq \lq dark\rq \rq\ GRB \citep[e.g.,][]{Jakobsson2004}. 
The dust extinction along the line of sight to the GRB is A$_{\rm V} \gtrsim$ 12 mag \citep{Perley2013a}.
The X-ray afterglow\footnote[1]{http://www.swift.ac.uk/xrt\_spectra/00279935/} detected by the {\it Swift} X-Ray Telescope (XRT) \citep{Guidorzi2007b} indicates the column density of $N_{\rm H}$=$(1.53^{+0.22}_{-0.20}) \times 10^{23}$ cm$^{-2}$ with a photon index of 1.86$^{+0.12}_{-0.12}$ \citep{Evans2007}. 
The X-ray photons are strongly obscured by the surrounding medium. 
The GRB position was localised by the {\it Swift}/XRT down to $\sim$ 1.5 arcsec radius. 
The host galaxy was identified within the XRT positional error circle \citep{Perley2013a}.
The redshift of the host galaxy, $z=2.0865$, is determined by H$\alpha$ emission \citep{Kruhler2015}.
The H$\alpha$ luminosity implies SFR$_{\rm H\alpha}$ = 26$^{+34}_{-17}$ $M_{\odot}$ yr$^{-1}$ \citep{Kruhler2015}. %Chabrier2003 IMF%
Observed host SED covering from rest-frame UV to near-infrared (NIR) wavelengths indicates SFR$_{\rm opt~SED}$ = 40.4 $^{+62.1}_{-3.0}$ $M_{\odot}$ yr$^{-1}$ with a stellar mass of $M_{*}=(3.08^{+1.89}_{-0.41})\times10^{10}$ $M_{\odot}$ \citep{Perley2013a}. %Chabrier2003 IMF%

Four years after the burst, \citet{Perley2013b} carried out Very Large Array wideband radio-continuum observations. 
Radio emission from the host galaxy was detected with S/N $\sim$ 3.0.
The radio SFR$_{\rm radio}$ is 817$\pm$300 $M_{\odot}$ yr$^{-1}$ calculated by a conversion formula by \citet{Murphy2011}, which is one order of magnitude higher than the optical estimate. %Kroupa IMF almost same as Chabrier2003%
This suggests a strongly obscured star-formation in the host galaxy. 
Or the radio detection is contaminated by the long-lasting radio afterglow even four years later.
So far, no FIR observations were conducted for the host galaxy.

\subsection{GRB 080207 host}
GRB 080207 host is one of the most explored dark GRB host galaxies at $z\sim2$.
After the GRB detection by {\it Swift}/BAT \citep{Racusin2008}, deep optical and NIR follow-up observations were performed \citep[e.g.,][]{Kuepcue2008,Cucchiara2008,Fugazza2008}.
No optical and NIR afterglows were detected.
The dust extinction along the line of sight to GRB 080207 is A$_{V} > 2.0$ mag \citep{Perley2013a}.
The X-ray afterglow was detected by the {\it Swift}/XRT and Chandra satellites, which localises the GRB with $\sim$0.5 arcsec accuracy \citep{Svensson2012}.
The X-ray afterglow\footnote[2]{http://www.swift.ac.uk/xrt\_spectra/00302728/} indicates $N_{\rm H}$=$(1.70^{+0.26}_{-0.24}) \times 10^{23}$ cm$^{-2}$ with a photon index of 2.2$^{+0.16}_{-0.15}$ \citep{Evans2007}.
The X-ray photons are strongly obscured by the surrounding medium. 
The extremely red host galaxy (R-K$>$5.4 AB mag.) was identified within the X-ray positional error circle \citep{Hunt2011,Svensson2012}.
The host redshift is 2.0858 determined by H$\alpha$ and [O~{\sc iii}]$\lambda 5007$ emission lines observed with the VLT/X-shooter \citep{Kruhler2012}.
The H$\alpha$ luminosity indicates SFR$_{\rm H\alpha}$=77$^{+86}_{-38}$ $M_{\odot}$ yr$^{-1}$ \citep{Kruhler2015}.
The SED fitting analysis of the host galaxy including rest-frame UV to NIR photometries implies SFR$_{\rm opt~SED}$=46.2$^{+271.9}_{-44.7}$ $M_{\odot}$ yr$^{-1}$ with a stellar mass of $M_{*}$=$(1.20^{+0.54}_{-0.48})\times10^{11}$ $M_{\odot}$ \citep{Perley2013a}.

The host galaxy was also detected by {\it Herschel}/PACS at rest-frame $\sim$ 30$\mu$m and 50$\mu$m \citep{Hunt2014}. 
Although the FIR peak of the SED is not covered by PACS detection, infrared SFR (SFR$_{\rm IR}$ = 170.1 $M_{\odot}$ yr $^{-1}$) suggests dust obscured star formation. %Chabrier2003 IMF%
The radio counterpart was detected with S/N $\sim$ 7 through the Karl G. Jansky Very Large Array (VLA) observations conducted three years after the burst \citep{Perley2013b}.
The radio SFR is 846$\pm$ 124 $M_{\odot}$ yr$^{-1}$, suggesting a intense star formation obscured by dust, although the contamination from radio afterglow can not be ruled out \citep{Perley2013b}. %Kroupa IMF almost same as Chabrier2003%
Here \citet{Perley2013b} derived the SFR$_{\rm radio}$ based on the formula by \citet{Murphy2011}.

Recently, several CO emissions were detected in the host galaxy. 
\citet{Arabsalmani2018} reported a CO(3-2) detection with  Plateau de Bure/NOEMA, suggesting a gas rich galaxy in contrast to other gas-poor GRB host galaxies \citep[e.g.,][]{Hatsukade2014}.
The molecular gas mass estimated from the CO(3-2) luminosity is $1.1\times 10^{11} M_{\odot}$.
The gas mass fraction of the host galaxy is $\sim0.5$, typical of star-forming galaxies with similar stellar mass and redshift.
In terms of the Kennicutt-Schmidt relation \citep{Kennicutt1998a}, i.e., SFR surface density as a function of molecular gas mass surface density, the host shares the same place as normal star-forming galaxies and submilimeter galaxies at $z=1-3$ in contrast to other GRB host galaxies \citep{Arabsalmani2018}. 
The CO(1-0), CO(2-1), and CO(4-3) emissions detected by IRAM 30m telescope, VLA, and ALMA also indicate the gas-rich property of the host sharing the similar molecular gas properties as other galaxies at the same redshift \citep{Michalowski2018,Hatsukade2019}.

The rest-frame FIR luminosity of the host is poorly constrained by {\it Herschel}/SPIRE \citep{Hunt2014}.
So far the host detection at the rest-frame FIR is not reported. 

\section{Observation and data reduction}
\label{observation}
ALMA cycle 5 observations (Project code: 2017.1.00337.S) were conducted for 070521 in band 9 on 23 May 2018, and 080207 in bands 8 and 9 on 12 and 19 May 2018, respectively. 
Each observation includes four spectral windows with bandwidths of 2 GHz composed of 128 channels (TDM mode).
We note that fine structure lines of [C~{\sc ii}]\,158\,$\mu$m and [N~{\sc ii}]\,205\,$\mu$m of the GRB 080207 host are covered by band 9 and 8, respectively, while the observed windows for the GRB 070521 host cover only rest-frame FIR continuum.

For the GRB 070521 observation, the central frequencies of the four spectral windows were 670.327, 675.035, 686.841, and 689.647 GHz.
The number of available antenna was 46 with the baseline lengths of 15.0$-$313.7m.
The exposure time was 17 min on source. 
J1517$-$2422 was used for bandpass and flux calibrations, while a gain calibrator was J1613+3412. %and J1653+3107.
The precipitable water vapour (PWV) was $\sim$0.35 mm during the observation.

For the GRB 080207 observations, the central frequencies were 471.537, 473.433 ([N~{\sc ii}]\,205$\mu$m), 459.862, and 461.482 GHz for band 8, and 613.780, 615.793 ([C~{\sc ii}]\,158$\mu$m), 628.675, 631.918 GHz for band 9.
The numbers of available antenna were 47 and 48 with the baseline lengths of 15.0$-$313.7 m and 15.0$-$455.6 m for band 8 and 9, respectively.
The on-source exposure times were 31 min in band 8 and 8 min in band 9.
In the band 8 observation, J1347+1217 was used for bandpass, flux, and gain calibrations.
In the band 9 observation, J1229+0203 was used for bandpass and flux calibrations, while a gain calibrator was J1332+0200.
The PWVs were $\sim$0.3 mm and $\sim$0.39 mm during the band 8 and 9 observations, respectively. 

The data were reduced with CASA version 5.1.1-5 for band 9 and version 5.4.0 for band 8 following standard data reduction scripts provided by Joint ALMA observatory.
Continuum frequency ranges were determined by excluding frequencies of atmospheric absorption lines and [C~{\sc ii}]\,158\,$\mu$m emission line.
The continuum frequency ranges were 669.327$-$671.327, 674.035$-$676.035, 685.841$-$687.841, and 688.647$-$690.647 GHz for the GRB 070521 observation in band 9.
The central frequency of the combined continuum map is 679.987 GHz.
The continuum frequency ranges for the GRB 080207 observations were 458.862$-$460.862, 460.482$-$462.482, 470.537$-$472.537, and 472.433$-$474.433 GHz in band 8, and 612.84$-$613.764, 614.277$-$614.688, 614.8$-$615.5, 616.0$-$616.7, 627.702$-$629.702, 631.2$-$632.0, and 632.5$-$632.7 GHz in band 9.
The central frequencies of the combined continuum maps are 466.6475 and 622.77 GHz in band 8 and 9, respectively.
Standard deviations of continuum images are 0.45, 0.10, and 1.4 mJy beam $^{-1}$ for the GRB 070521 host at 679.987 GHz, the GRB 080207 host at 466.6475 GHz, and the GRB 080207 host at 622.77 GHz, respectively.
The standard deviations were measured by avoiding the source positions and edges of the images.
\lq \lq Natural\rq \rq\ weight was adopted to create CLEAN images with spectral velocity bins of 50 km s$^{-1}$.
Beam sizes of the band 9 observation for the GRB 070521 host were 590 and 450 mas for major and minor axes, respectively.
Major and minor beam sizes of the GRB 080207 host observations were 940 and 550 mas in band 8 and 510 and 440 mas in band 9, respectively.
The observational configurations are summarised in Table \ref{tab1}.

\begin{table*}
	\centering
	\caption{
	Observational configurations.
    }
	\label{tab1}
	\begin{flushleft}
	\begin{tabular}{|l|c|c|c|}\hline
	         & 070521 & 080207 ([C~{\sc ii}]/band 9) & 080207 ([N~{\sc ii}]/band 8)\\ \hline \relax
	         & ($z^{\rm opt}_{\rm spec}$=2.0865$^{a}$) & \multicolumn{2}{|c|}{($z^{\rm opt}_{\rm spec}$=2.0858$^{b}$)} \\ \hline \relax
    ALMA band     & 9      & 9 & 8 \\ \relax
    Number of antenna & 46 & 48 & 47 \\ \relax
    Spectral window 1 (GHz)& 669.327$-$671.327& 612.780$-$614.780 & 470.537$-$472.537 \\ \relax
    Spectral window 2 (GHz)& 674.035$-$676.035& 614.793$-$616.793 ([C~{\sc ii}]\,158$\mu$m) & 472.433$-$474.433 ([N~{\sc ii}]\,205$\mu$m) \\ \relax
    Spectral window 3 (GHz)& 685.841$-$687.841& 627.675$-$629.675 & 458.862$-$460.862 \\ \relax
    Spectral window 4 (GHz)& 688.647$-$690.647& 630.918$-$632.918 & 460.482$-$462.482 \\ \relax
    Rest-frame wavelength 1 ($\mu$m) & 144.684$-$145.116 & 158.028$-$158.543 & 205.597$-$206.471 \\ \relax
    Rest-frame wavelength 2 ($\mu$m) & 143.676$-$144.103 & 157.512$-$158.024 ([C~{\sc ii}]\,158$\mu$m) & 204.776$-$205.642 ([N~{\sc ii}]\,205$\mu$m) \\ \relax
    Rest-frame wavelength 3 ($\mu$m) & 141.210$-$141.622 & 154.290$-$154.781 & 210.806$-$211.724 \\ \relax
    Rest-frame wavelength 4 ($\mu$m) & 140.637$-$141.045 & 153.499$-$153.986 & 210.067$-$210.980 \\ \relax
    Exp. time on source (min) & 17 & 5 & 31 \\ \relax
    Beam major/minor axis (mas)& 590/450 & 510/440 & 940/550 \\ \relax
    Beam position angle (deg)& $-$15.93 & $-$76.53 & $-$60.03 \\ \hline 
    \end{tabular}\\
    $^{a}$Measured by H$\alpha$ \citep{Kruhler2015}.
    $^{b}$Measured by H$\alpha$ and [O~{\sc iii}]$\lambda 5007$ \citep{Kruhler2012}. \\
    \end{flushleft}
\end{table*}

\begin{figure*}
	\includegraphics[width=2.25in]{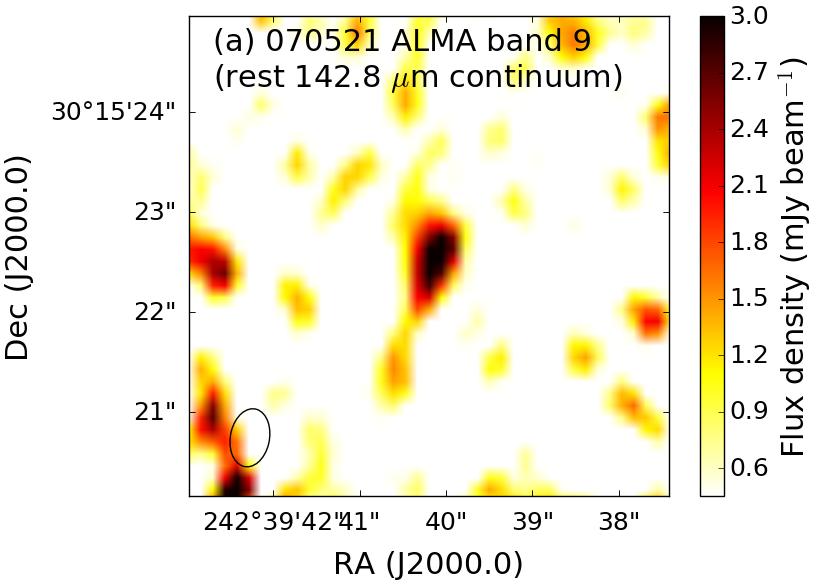}
	\includegraphics[width=2.25in]{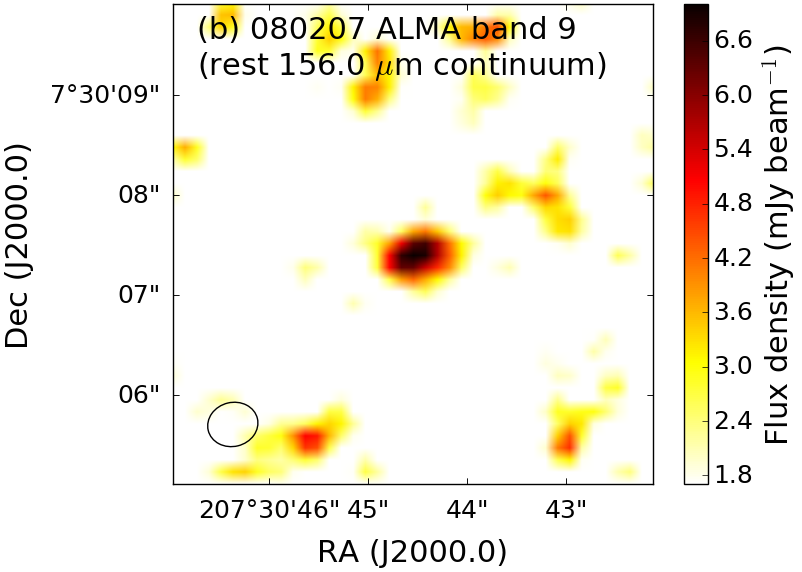}
	\includegraphics[width=2.25in]{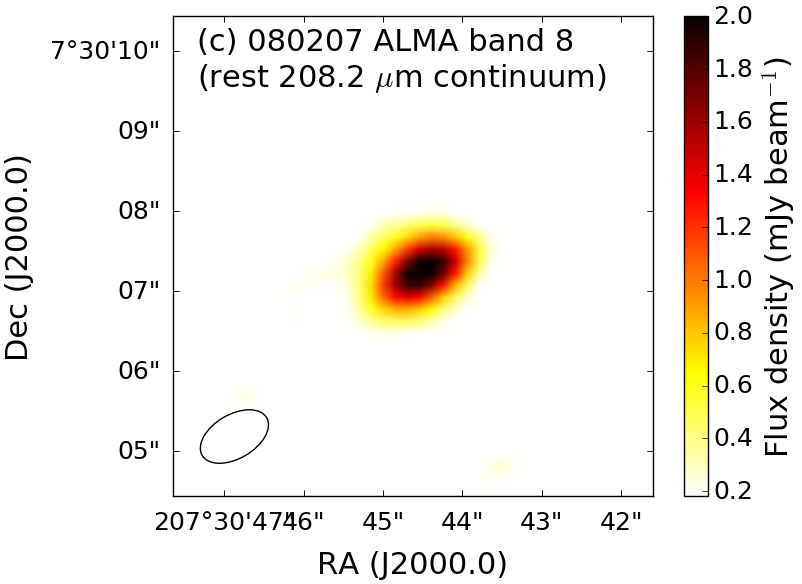}
    \caption{
    Rest-frame FIR continuum images obtained with (left) ALMA band 9 for the GRB 070521 host, (middle) band 9 for the GRB 080207 host, and (right) band 8 for the GRB 080207 host. 
    Beam sizes are indicated by ellipses.
    Colour ranges are between $1$ $\sigma$ noise levels and peak flux densities.
    }
    \label{fig1}
\end{figure*}

\begin{figure}
	\includegraphics[width=3.45in]{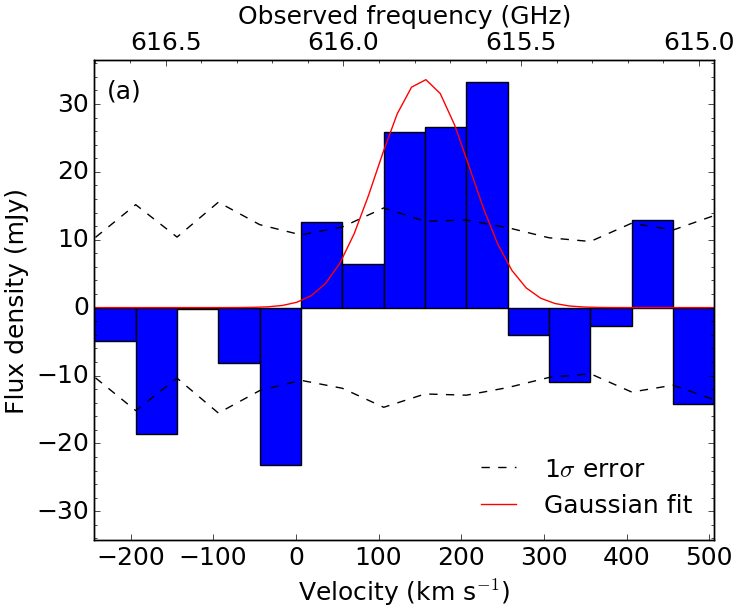}
	\includegraphics[width=3.45in]{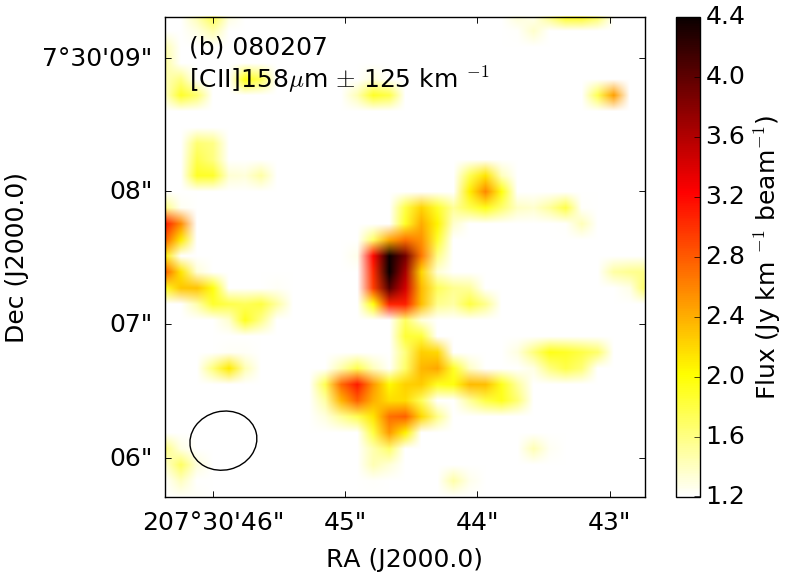}
    \caption{
    (Top) [C~{\sc ii}]\,158\,$\mu$m spectrum with a spectral resolution of 50 km s$^{-1}$ at the position of the GRB 080207 host galaxy.
    The velocity axis is relative to $z=2.0858$ measured by [O~{\sc iii}]$\lambda 5007$ and H$\alpha$ \citep{Kruhler2012}.
    The continuum emission is subtracted from the spectrum.
    Noise levels, $\pm$1$\sigma$, are shown by dashed lines.
    Red solid line is the best-fit Gaussian function with a central position at 155 km s$^{-1}$.
    (Bottom) [C~{\sc ii}]\,158\,$\mu$m map of the GRB 080207 host.
    The [C~{\sc ii}]\,158\,$\mu$m emission is integrated over $\pm$125 km s$^{-1}$ after subtraction of the continuum emission.
    Beam size is indicated by an ellipse.
    Colour range is between 1 $\sigma$ noise level and peak flux density.
    The apparent size of the velocity-integrated [C~{\sc ii}]\,158$\mu$m emission is consistent with a point source.
    }
    \label{fig2}
\end{figure}

We detected FIR continuum fluxes at the positions of the host galaxies with S/N$\sim$8 for 070521 in band 9, and S/N$\sim$19 and 7 for 080207 in band 8 and 9, respectively (Fig. \ref{fig1}).
CASA {\it imfit} task was utilised to fit continuum images and to estimate the fluxes and the errors.
[C~{\sc ii}]\,158\,$\mu$m emission of the GRB 080207 host was marginally detected around the expected frequency for $z=2.0858$ \citep{Kruhler2012} at the position of the GRB 080207 host by removing the continuum emission (Fig. \ref{fig2}a).
The velocity offset between [C~{\sc ii}]\,158$\mu$m and optical emission lines is 155 km s$^{-1}$ measured by a Gaussian fitting.
The CO(1-0) velocity of the GRB 080207 host galaxy also shows an offset from the optical emission lines \citep{Hatsukade2019}.
The CO(1-0) velocity offset is $\sim100$ km s$^{-1}$ which is comparable to the [C~{\sc ii}]\,158$\mu$m velocity offset.
The [C~{\sc ii}]\,158$\mu$m spectrum was integrated over $\pm$ 125 km s$^{-1}$ velocity width to estimate the emission line flux.
This velocity integration was performed after the 50 km s$^{-1}$ velocity binning in the CLEAN process mentioned above.
The velocity-integrated [C~{\sc ii}]\,158$\mu$m emission is located at the position of the GRB 080207 host (Fig. \ref{fig2}b). 
The {\it imfit} task indicated that the apparent size of [C~{\sc ii}]\,158$\mu$m on sky is consistent with a point source.
The S/N of the [C~{\sc ii}]\,158$\mu$m emission line is $\sim$4 after the velocity integration.
We summarise the observational results in Table \ref{tab2}.

\begin{table*}
	\centering
	\caption{
	Measured values.
    }
	\label{tab2}
	\begin{flushleft}
	\begin{tabular}{|l|c|c|c|}\hline
	         & 070521 & 080207 ([C~{\sc ii}]/band 9) & 080207 ([N~{\sc ii}]/band 8) \\ \hline \relax
    FIR continuum flux (mJy) & 3.58$\pm$0.45 & 11.9$\pm$1.7 & 3.37$\pm$0.18\\ \relax
    FIR continuum convolved size (mas)$^{a}$& 828$\pm$126/427$\pm$37 & 738$\pm$83/488$\pm$40 & 1119$\pm$46/732$\pm$22 \\ \relax
    FIR continuum deconvolved size (mas)$^{a}$& point source & 533$\pm$123/201$\pm$193 & 606$\pm$92/476$\pm$62 \\ \relax
    FIR continuum deconvolved physical size (kpc)$^{a}$ & $<$7.1$^{c}$ & 4.55$\pm$1.05/1.72$\pm$1.65 & 5.18$\pm$0.79/4.06$\pm$0.53 \\ \relax 
    FIR continuum position angle (deg)$^{b}$&168.2$\pm$4.9&97.6$\pm$7.8& 118.0$\pm$2.7\\ \relax
    Emission-line flux (Jy km s$^{-1}$)& - & 4.4$\pm$1.2$^{d}$ &$<$0.28$^{e}$\\ \relax
    [C~{\sc ii}]\,158$\mu$m offset velocity (km s$^{-1}$)& - & 155$^{f}$ & - \\ \relax
    [C~{\sc ii}]\,158$\mu$m FWHM (km s$^{-1}$)& - & 132 & - \\ \hline \relax
    \end{tabular}\\
    $^{a}$Major/minor axes FWHM.
    $^{b}$Measured for the convolved image.
    $^{c}$Upper limit measured from the major axis of the convolved image.
    $^{d}$Integrated over $\pm$ 125 km s$^{-1}$ and integrated spatially.
    $^{e}$3 $\sigma$ upper limit assuming the velocity width of [C~{\sc~ii}]\,158$\mu$m.
    $^{f}$Relative to $z=2.0858$ measured by [O~{\sc iii}]$\lambda 5007$ and H$\alpha$ \citep{Kruhler2012}.
    \end{flushleft}
\end{table*}

\section{SED Modelling and results}
\label{analysis_results}
The rest-frame FIR continuum emissions were detected for the GRB 070521 and 080207 host galaxies by ALMA (see Section \ref{observation}). 
The FIR flux measurements allow us to calculate obscured SFR accurately.
Dust emission was modelled by SED fitting analysis with Multi-wavelength Analysis of Galaxy Physical Properties: MAGPHYS \citep{Cunha2008}, high-z version \citep{Cunha2015}, which is computed based on the energy balance.
The energy budget determined by dust obscuration from the UV to NIR range is re-allocated in the mid-infrared (MIR) to FIR range using empirically calibrated dust emission: PAHs and (stochastically heated)  hot dust in the MIR and two components (warm and cold) of dust in thermal equilibrium.

Based on the modelling of the stellar emission by \citet{Bruzual2003}, which adopted the single-star IMF, and dust attenuation model of \citet{Charlot2000}, the MAGPHYS uses the adjustable parameters concerning star formation history such as timescale ($\gamma$), age ($t_{gal}$), amplitude of random star bursts (A), and  stellar metallicity ($Z_{\rm star}$) as well as total $V$-band optical depth of the dust seen by young stars in their birth clouds ($\hat{\tau_{V}}$), and fraction of $\hat{\tau_{V}}$ contributed by dust in the diffuse interstellar medium ($\mu$) \citep[e.g,][]{Cunha2008,Cunha2015,Hunt2019}.
To model the infrared SED, the contributions of dust emission both from stellar birth clouds (SB) and from ambient inter-stellar medium (ISM) are considered. 
For the SB, the fractional contribution by PAHs emission ($\xi^{\rm BC}_{\rm PAH}$), hot MIR emission ($\xi^{\rm BC}_{\rm MIR}$), and warm dust emission ($\xi^{\rm BC}_{\rm W}$) are modelled.  
For the ambient ISM, contribution from cold($\xi^{\rm ISM}_{\rm C}$) dust in the equilibrium temperature are added. 
And the fractional contribution of cold dust ($f_{\mu}$) to the total dust luminosity ($L_{d}^{tot} = L_{d}^{BC}+L_{d}^{ISM}$) is defined in order to describe all the contribution of different components to total dust luminosity. 
These two parts (stellar emission with dust attenuation and infrared SED) are connected together, so it is important to get (at least one) photometric data point  in the FIR range to constrain the modelling by this energy balance principle.

In this work, the photometries at the wavelength longer than $\sim$500$\mu$m were not used in order to avoid the possible contamination from the long-lived afterglows \citep{Perley2013b} in the modelling.
The IMF is adopted from \citet{Chabrier2003}. 
The multi-wavelength data of the GRB 070521 and 080207 hosts are summarised in Tables \ref{tab3} and \ref{tab4}.

\begin{table*}
	\centering
	\caption{
	Multi-wavelength data of GRB 070521 host galaxy.
    }
	\label{tab3}
	\begin{flushleft}
	\begin{tabular}{|l|c|c|c|c|}\hline
	\multicolumn{5}{|c|}{070521}  \\ \hline \relax
    Observed wavelength ($\mu$m) & Band & Flux ($\mu$Jy) & Telescope/Instrument & Reference \\ \hline \relax
%    0.43221$^{a}$&B&0.04$\pm$0.01&Keck/LRIS & \citet{Perley2013a} \\ \relax
%    0.54037$^{a}$&V&0.1$\pm$0.03&Keck/LRIS & \citet{Perley2013a} \\ \relax
%    0.75497$^{a}$&I&0.28$\pm$0.11&Keck/LRIS & \citet{Perley2013a} \\ \relax
%    0.90940$^{a}$&z&0.26$\pm$0.09&Keck/LRIS & \citet{Perley2013a} \\ \relax
%    1.24161$^{a}$&J&1.59$\pm$0.58&Keck/NIRC & \citet{Perley2013a} \\ \relax
%    1.58513$^{a}$&H&2.23$\pm$0.21&Keck/NIRC & \citet{Perley2013a} \\ \relax
%    2.18186$^{a}$&K&3.02$\pm$0.55&Keck/NIRC & \citet{Perley2013a} \\ \relax
%    3.50751$^{a}$&3.6um&6.55$\pm$0.83&Spitzer/IRAC & \citet{Perley2013a} \\ \relax
%    4.43658$^{a}$&4.5um&9.04$\pm$0.96&Spitzer/IRAC & \citet{Perley2013a} \\ \relax
    0.43221$^{a}$&B&0.04$\pm$0.01&Keck/LRIS & \citet{Perley2013a} \\ \relax
    0.54037$^{a}$&V&0.1$\pm$0.03&Keck/LRIS & \citet{Perley2013a} \\ \relax
    0.75497$^{a}$&I&0.28$\pm$0.11&Keck/LRIS & \citet{Perley2013a} \\ \relax
    0.90940$^{a}$&z&0.26$\pm$0.09&Keck/LRIS & \citet{Perley2013a} \\ \relax
    1.2416$^{a}$&J&1.59$\pm$0.58&Keck/NIRC & \citet{Perley2013a} \\ \relax
    1.5851$^{a}$&H&2.23$\pm$0.21&Keck/NIRC & \citet{Perley2013a} \\ \relax
    2.1818$^{a}$&K&3.02$\pm$0.55&Keck/NIRC & \citet{Perley2013a} \\ \relax
    3.5075$^{a}$&3.6um&6.55$\pm$0.83&Spitzer/IRAC & \citet{Perley2013a} \\ \relax
    4.4365$^{a}$&4.5um&9.04$\pm$0.96&Spitzer/IRAC & \citet{Perley2013a} \\ \relax
    440.87$^{b}$&band9&3580$\pm$448&ALMA/band9 & This work \\ \relax
    57322&5.23GHz&28$\pm$10.3$^{c}$&VLA & \citet{Perley2013b} \\ \hline 
    \end{tabular}\\
    $^{a}$Effective wavelength (http://svo2.cab.inta-csic.es/theory/fps3/index.php?mode=browse).\\
    $^{b}$Central wavelength of the spectral windows used for the continuum image (Fig. \ref{fig1}a).\\
    $^{c}$NOT used in the SED fitting analysis to avoid the possible contaminated flux from the long-lived afterglow.
    \end{flushleft}
\end{table*}

\begin{table*}
	\centering
	\caption{
	Multi-wavelength data of GRB 080207 host galaxy.
    }
	\label{tab4}
	\begin{flushleft}
	\begin{tabular}{|l|c|c|c|c|}\hline
	\multicolumn{5}{|c|}{080207}  \\ \hline \relax
    Observed wavelength ($\mu$m) & Band & Flux ($\mu$Jy) & Telescope/Instrument & Reference \\ \hline \relax
    0.47067$^{a}$&g&0.04$\pm$0.01&Keck/LRIS & \citet{Svensson2012} \\ 
    0.63757$^{a}$&R&0.093$\pm$0.026&VLT/VIMOS & \citet{Hunt2011} \\ 
    0.75497$^{a}$&I&0.17$\pm$0.05&Keck/LRIS & \citet{Svensson2012} \\ 
    0.95716$^{a}$&z$^{\prime}$&0.35$\pm$0.06&Gemini/GMOS & \citet{Hunt2011} \\ 
    1.1029$^{a}$&F110W&1.75$\pm$0.17&HST/WFC3 & \citet{Svensson2012} \\ 
    1.2&J&1.6$\pm$0.3&VLT/SINFONI & \citet{Hunt2011} \\ 
    1.5785$^{a}$&F160W&2.9$\pm$0.7&HST/NICMOS (NIC3) & \citet{Hunt2011} \\ 
    1.5785$^{a}$&F160W&2.27$\pm$0.34&HST/NICMOS (NIC3) & \citet{Svensson2012} \\ 
    2.1063$^{a}$&K$^{\prime}$&6.25$\pm$1.62&Gemini/NIRI & \citet{Svensson2012} \\ 
    2.1521$^{a}$&Ks&7.3$\pm$1.0&VLT/ISAAC & \citet{Hunt2011} \\ 
    3.5075$^{a}$&3.6um&14.40$\pm$0.31&Spitzer/IRAC & \citet{Hunt2011} \\ 
    4.4365$^{a}$&4.5um&15.51$\pm$0.44&Spitzer/IRAC & \citet{Hunt2011} \\ 
    5.6281$^{a}$&5.7um&18.53$\pm$1.58&Spitzer/IRAC & \citet{Hunt2011} \\ 
    7.5891$^{a}$&7.9um&12.52$\pm$1.76&Spitzer/IRAC & \citet{Hunt2011} \\ 
    23.680$^{a}$&24um&92.43$\pm$6.50&Spitzer/MIPS & \citet{Hunt2011} \\ 
    97.903$^{a}$&100um&2200$\pm$600&{\it Herschel}/PACS & \citet{Hunt2014} \\ 
    153.94$^{a}$&160um&5900$\pm$1400&{\it Herschel}/PACS & \citet{Hunt2014} \\ 
    242.82$^{a}$&250um&$<$19500$^{c}$&{\it Herschel}/SPIRE & \citet{Hunt2014} \\ 
    340.89$^{a}$&350um&$<$20400$^{c}$&{\it Herschel}/SPIRE & \citet{Hunt2014} \\ 
    450&450um&$<$52483$^{c}$&JCMT/SCUBA2 & \citet{Svensson2012} \\ 
    481.38$^{b}$&band9&11900.0$\pm$1700.0&ALMA/band9 & This work \\ 
    482.25$^{a}$&500um&$<$21900$^{c}$&{\it Herschel}/SPIRE & \citet{Svensson2012} \\ 
    642.43$^{b}$&band8&3370$\pm$180$^{d}$&ALMA/band8 & This work \\ 
    850&850um&$<$13183$^{c}$&JCMT/SCUBA2 & \citet{Svensson2012} \\ 
    2103.4&band4&123.0$\pm$24$^{d}$&ALMA/band4 & \citet{Hatsukade2019} \\ 
    8137.7&36.83GHz&$<$22.4$^{c}$&VLA & \citet{Svensson2012} \\ 
    57322&5.23GHz&17.1$\pm$2.5$^{d}$&VLA & \citet{Perley2013b} \\ \hline
    \end{tabular}\\
    $^{a}$Effective wavelength (http://svo2.cab.inta-csic.es/theory/fps3/index.php?mode=browse).\\
    $^{b}$Central wavelengths of the spectral windows used for the continuum images (Fig. \ref{fig1}b and c).\\
    $^{c}$3$\sigma$ upper limit, which is not taken into account in the SED fitting analysis.\\ 
    $^{d}$NOT used in the SED fitting analysis to avoid the possible contaminated flux from the long-lived afterglow.
    \end{flushleft}
\end{table*}

\begin{figure*}
	\includegraphics[width=3.45in]{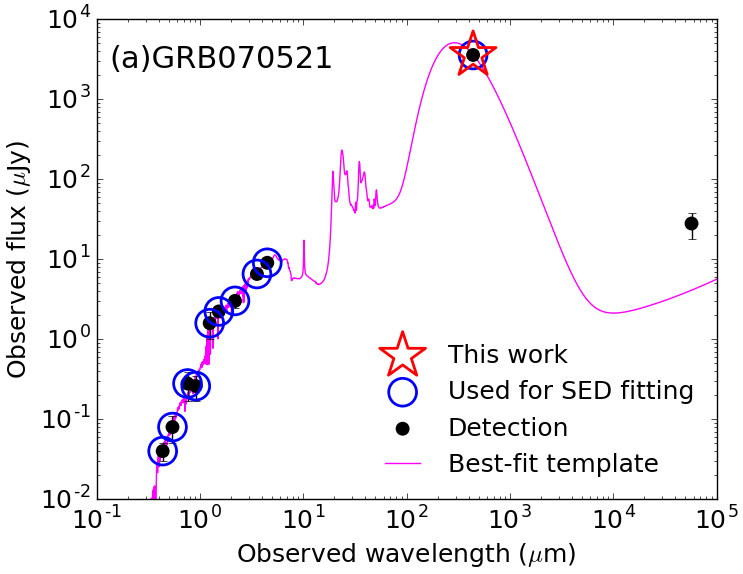}
	\includegraphics[width=3.45in]{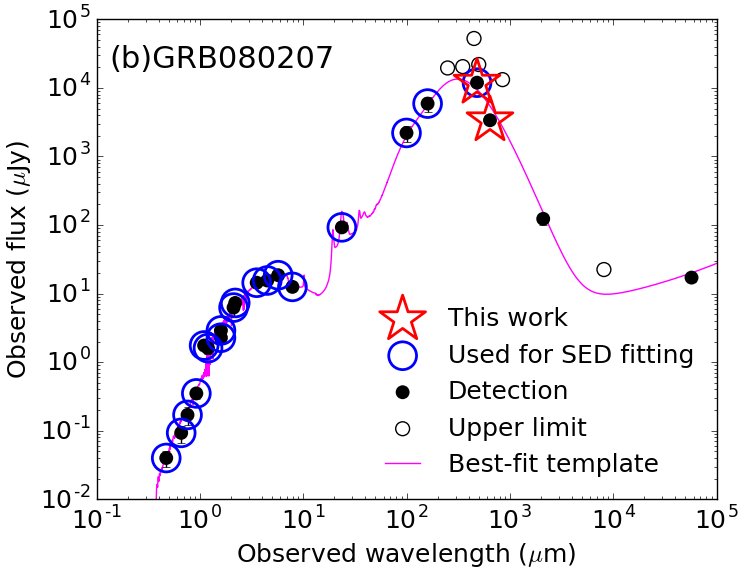}
    \caption{
    Spectral energy distributions of the GRB 070521 (left) and 080207 (right) host galaxies (black dots).
    ALMA photometries measured in this work are marked by red stars.
    Photometies at wavelengths longer than $\sim$500$\mu$m were excluded from the SED fitting analysis to avoid the possible contaminated flux from the long-lived afterglows.
    Photometries used for the SED fitting analysis are marked by blue circles.
    The best fit results of the SED fitting analysis with MAGPHYS HIGHZ \citep{Cunha2008,Cunha2015} is shown with magenta curves. 
    3$\sigma$ upper limits are demonstrated by black circles. 
    Details of the photometries are summarised in Tables \ref{tab3} and \ref{tab4}.
    }
    \label{fig3}
\end{figure*}

The best fit SED models are shown in Fig. \ref{fig3}.
The observed SEDs of the GRB 070521 and 080207 hosts are successfully reproduced by the best fit models except for the radio flux of the GRB 070521 host.
%\textcolor{red}{Note that we used only photometries up to $\sim$500$\mu$m to avoid the possible contamination from emissions of long-lived afterglows.}
Physical parameters derived from the best fit templates are summarised in Table \ref{tab5}.
We use a \lq \lq total infrared luminosity\rq \rq, $L_{\rm TIR}$, defined as an integration of the bolometric dust thermal emissions originating from the stellar birth clouds and the ambient ISM \citep{Cunha2008,Cunha2015}.
$L_{\rm TIR}$ was converted to the FIR luminosity, $L_{\rm FIR}$, by assuming a conversion factor, log ($L_{\rm TIR}$/$L_{\rm FIR}$)=0.24 \citep{Calzetti2000}.
Here $L_{\rm FIR}$ is defined as a luminosity integrated between rest-frame 40 and 120 $\mu$m.
We used the deconvolved physical size of the GRB 080207 host measured in band 8 to calculate the TIR and FIR surface densities, $\Sigma_{\rm TIR}$ and $\Sigma_{\rm FIR}$, because the S/N of the continuum flux in band 8 is higher than that in band 9.
The $\Sigma_{\rm TIR}$ and $\Sigma_{\rm FIR}$ were calculated as $L_{\rm TIR}$/($\pi r_{\rm IR}^{2}$) and $L_{\rm FIR}$/($\pi r_{\rm IR}^{2}$), respectively, where $r_{\rm IR}$ is the half of the deconvolved physical size (FWHM/2.0) in major axis.
Only lower limits on $\Sigma_{\rm TIR}$ and $\Sigma_{\rm FIR}$ are available for the GRB 070521 host, because the FIR image is not spatially resolved by the {\it imfit} task.

%\textcolor{magenta}{XXX: 070521 has only one data at FIR. Why is it possible to constrain dust temperature with a so accurate dust mass? Is this related with large uncertainty of SFR? I need answer.}

In Table \ref{tab5}, uncertainties of dust properties of the GRB 070521 host, i.e., dust mass and dust temperature errors, are probably underestimated by MAGPHYS.
There is only one photometric data point at the rest-frame FIR wavelength. 
In general, a degeneracy between dust mass and dust temperature can not be properly solved from single photometric data in FIR. 
Therefore the dust properties of the GRB 070521 host are likely more uncertain.
It is possible that the derived dust properties significantly change when additional FIR data are included in the SED fitting analysis in the future.

\begin{table}
	\centering
	\caption{
	Physical parameters of GRB hosts.
    }
	\label{tab5}
	\begin{flushleft}
	\begin{tabular}{|l|c|c|}\hline
	         & 070521 & 080207 \\ \hline \relax
    Stellar mass ($M_{\odot}$) & (4.49$^{+2.12}_{-0.02}$)$\times 10^{10}$ & (1.70$^{+0.08}_{-0.35}$)$\times 10^{11}$ \\ \relax
    SFR ($M_{\odot}$ yr$^{-1}$) & 49.85$^{+72.33}_{-2.86}$ & 123.40$^{+25.19}_{-21.78}$ \\ \relax
    Total IR luminosity ($L_{\odot}$)& (6.09$^{+8.03}_{-0.96}$) $\times 10^{11}$ & (1.82$^{+0.23}_{-0.23}$) $\times 10^{12}$  \\ \relax
    FIR luminosity ($L_{\odot}$)$^{a}$ & (3.50$^{+4.62}_{-0.55}) \times 10^{11}$ & $(1.05^{+0.13}_{-0.13}) \times 10^{12}$ \\ \relax
    $\Sigma_{\rm TIR}$ ($L_{\odot}$ kpc$^{-2}$)$^{b}$ & $>1.5\times 10^{10}$ &$(8.65^{+4.89}_{-2.95}) \times 10^{10}$ \\ \relax
    $\Sigma_{\rm FIR}$ ($L_{\odot}$ kpc$^{-2}$)$^{ab}$ & $>8.8\times 10^{9}$ &$(4.98^{+2.81}_{-1.70}) \times 10^{10}$ \\ \relax
    Dust mass ($M_{\odot}$)& (1.52$^{+1.34}_{-0.76}$) $\times 10^{8}$ & (5.45$^{+3.61}_{-1.80}$) $\times 10^{8}$  \\ \relax
    Dust temperature ($K$)$^{c}$& 31.89$^{+19.36}_{-2.66}$ & 39.86$^{+1.10}_{-4.11}$ \\ \relax 
    [C~{\sc ii}]\,158\,$\mu$m luminosity ($L_{\odot}$)& - & (7.85$\pm 2.14$)$\times 10^{8}$ \\ \relax 
    $Z_{\rm PP04~N2}$ ($Z_{\odot}$)$^{d}$ & - & 0.69$^{+0.29}_{-0.20}$ \\ \relax
    $Z_{\rm [NII]205\mu m/[CII]158\mu m}$ ($Z_{\odot}$)$^{e}$ & - & $\lesssim$1 \\ \hline
    \end{tabular}\\
    $^{a}$A conversion factor by \citet{Calzetti2000}, log($L_{\rm TIR}$/$L_{\rm FIR}$)=0.24, is assumed.
    $^{b}\Sigma_{\rm IR}$ = $L_{\rm IR}/(\pi r_{\rm IR}^{2})$. To calculate $r_{\rm IR}$ of the GRB 080207 host, we used the deconvolved FWHM/2.0 of the major axis in band 8 due to the higher S/N than band 9.
    $^{c}$Averaged temperature of the two dust components with luminosity weights.
    $^{d}$Optical emission-line diagnostics \citep{Kruhler2015} based on PP04~N2 method \citep{Pettini2004} normalised by the Solar value of 12+log(O/H)=8.69 \citep{Asplund2009}.
    $^{e}$FIR emission-line diagnostics \citep{Nagao2012} assuming log$n_{\rm e}$=3 cm$^{-1}$ and log$U$=$-$2.5, where $n_{\rm e}$ and $U$ are electron density and ionisation parameter.
    \end{flushleft}
\end{table}

\section{Discussion}
\label{discussion}
\subsection{Comparison of physical properties in previous studies}
\label{comparison}
Here we compare our results with physical parameters of GRB 070521 and 080207 hosts derived in previous studies.
We focus on SFR and dust properties, since these two parameters are updated by our ALMA observations.

\subsubsection{GRB 070521 host}
By adding rest-frame FIR photometry, we calculated the SFR of the host from SED fitting analysis (see Section \ref{analysis_results}).
The derived SFR of the GRB 070521 host is 49.9$^{+72.3}_{-2.9}$ $M_{\odot}$ yr$^{-1}$.
The extinction-corrected optical SFRs in previous studies are SFR$_{\rm H\alpha}=26^{+34}_{-17}$ $M_{\odot}$ yr$^{-1}$ and SFR$_{\rm opt~SED}=40.4^{+62.1}_{-3.0}$ $M_{\odot}$ yr$^{-1}$ \citep{Perley2013a,Kruhler2015}.
Our SFR estimate is consistent with previous optical measurements within errors, while this work added the rest-frame FIR photometric data.
The host galaxy does not show a strong dust-obscured star formation.
The extremely high SFR in radio (SFR$_{\rm radio}$=817$\pm$300 $M_{\odot}$ yr$^{-1}$) is probably due to the contribution from the long-lived afterglow as cautioned by \citet{Perley2013b}.
The radio flux four years after the burst is 28.0 $\mu$Jy at 5.23 GHz \citep{Perley2013b}, which corresponds to $\sim$3$-$100 $\mu$Jy at 600 GHz assuming that the radio flux originates from the afterglow with typical spectral slopes, $\alpha$=$-$1/2 to 1/3, \citep{Sari1998}.
These values are negligible compared with the FIR flux detected by ALMA, i.e., 11.9 mJy.
Therefore the afterglow contamination to the derived SFR$_{\rm IR}$ is negligible.

In Fig. \ref{fig3}a, there is an obvious radio excess from the best-fit template.
This could be because of an active galactic nuclei (AGN) rather than the afterglow.
The optical emission-line diagnostics, the Baldwin-Phillips-Terlevich (BPT) diagram \citep{Baldwin1981}, is not available for this host galaxy because the only H$\alpha$ emission was detected with VLT/X-shooter \citep{Kruhler2015}. 
Other emission lines including H$\beta$, [O~{\sc iii}]$\lambda$5007, and [N~{\sc ii}]$\lambda$6584 were covered by the VLT/X-shooter observation, but not detected \citep{Kruhler2015}. 
The 3 sigma upper limit on log([N~{\sc ii}]$\lambda$6584/H$\alpha$) including the H$\alpha$ flux uncertainty is -0.27. 
This value is close to the boundary between star-forming galaxies and AGNs \citep[e.g.,][]{Baldwin1981}.
Therefore, the existence of the AGN is ambiguous from the BPT diagram.

The derived stellar mass of $(4.49^{+2.12}_{-0.02})$ $\times 10^{10}$ $M_{\odot}$ is consistent with that in the previous study, $M_{*}$=$(3.08^{+1.89}_{-0.41})\times 10^{10}$ $M_{\odot}$ \citep{Perley2013a}.
As mentioned in Section \ref{analysis_results}, the dust properties are likely more uncertain than the errors derived by MAGPHYS, because there is only one photometric data in the rest-frame FIR.
Multi-band FIR photometries are necessary to constrain dust properties in a more reliable way.

\subsubsection{GRB 080207 host}
The derived SFR of the GRB 080207 host is 123.4$^{+25.19}_{-21.78}$ $M_{\odot}$ yr$^{-1}$.
\citet{Hunt2014} estimated the SFR$_{\rm IR}$ of 170.1 $M_{\odot}$ yr$^{-1}$ by including {\it Herschel}/PACS detections.
Although the SFR$_{\rm IR}$ error is not explicitly provided in the literature, the S/N $\sim$4 of PACS detection would roughly correspond to $\sim$25\% uncertainty in SFR$_{\rm IR}$.
Supposing this error budget, our measurement is consistent with the SFR$_{\rm IR}$ by \citet{Hunt2014} within the observational error.
The derived SFR$_{\rm IR}$ is significantly lower than that of the radio observation, SFR$_{\rm radio}$=846$\pm$124 $M_{\odot}$ yr$^{-1}$ \citep{Perley2013b}.
The radio flux three years after the burst is 17.1 $\mu$Jy at 5.23 GHz \citep{Perley2013b}, which corresponds to $\sim$2$-$70 $\mu$Jy at 600 GHz assuming that the radio flux originates from the afterglow with typical spectral slopes \citep{Sari1998}.
These values are negligible compared with the FIR flux detected by ALMA, i.e., 3.58 mJy.
Therefore the afterglow contamination to the derived SFR$_{\rm IR}$ is negligible.

Both of our best-fit SED model (Fig. \ref{fig3}b) and that of \citet{Hunt2014} are consistent with the radio detection, but both result in much lower SFRs than the radio.
This probably means that the high SFR$_{\rm radio}$ is because of a too high flux-SFR conversion factor used for this host galaxy, rather than the afterglow/AGN contamination.

The averaged dust temperature in our analysis is $T_{\rm dust}$=39.36$^{+1.10}_{-4.11}$ $K$. 
\citet{Hunt2014} derived $T_{\rm dust}=61.3$ $K$ based on {\it Herschel} data with GRASIL \citep{Silva1998} that has dust components at a range of temperatures.
\citet{Hatsukade2019} performed a modified blackbody fit to FIR photometries by adding ALMA data, and derived $T_{\rm dust}$=37$\pm4$ $K$.
Our estimate is more consistent with \citet{Hatsukade2019}. 
The difference from \citet{Hunt2014} is probably due to the additional data from ALMA.
The dust mass derived by our analysis is $(5.45^{+3.61}_{-1.80})\times 10^{8}$ $M_{\odot}$.
\citet{Hunt2014} and \citet{Hatsukade2019} estimated dust mass of 1.4$\times 10^{8}$ $M_{\odot}$ and 1.5$\times 10^{8}$ $M_{\odot}$, respectively.
Our estimate is systematically larger than these values.
The difference is likely due to the new data by ALMA band 9 covering the peak FIR emission from dust and different method to fit IR data.
\citet{Hatsukade2019} fitted a modified black body with single temperature to the IR photometries.
We assumed two components of dust temperatures in the MAGPHYS SED fitting analysis.
This assumption traces wider range of dust temperatures than the single component, resulting in a larger dust mass integrated over multi temperatures.

The previous stellar mass estimates are 1.48$\times 10^{11}$ $M_{\odot}$ \citep{Hunt2014} and $(1.20^{+0.54}_{-0.48}) \times 10^{11}$ $M_{\odot}$ \citep{Perley2013a}.
The derived stellar mass of $(1.70^{+0.08}_{-0.35})\times 10^{11}$ $M_{\odot}$ in this work is consistent with the literature.

\subsection{Main sequence}
\begin{figure}
	\includegraphics[width=\columnwidth]{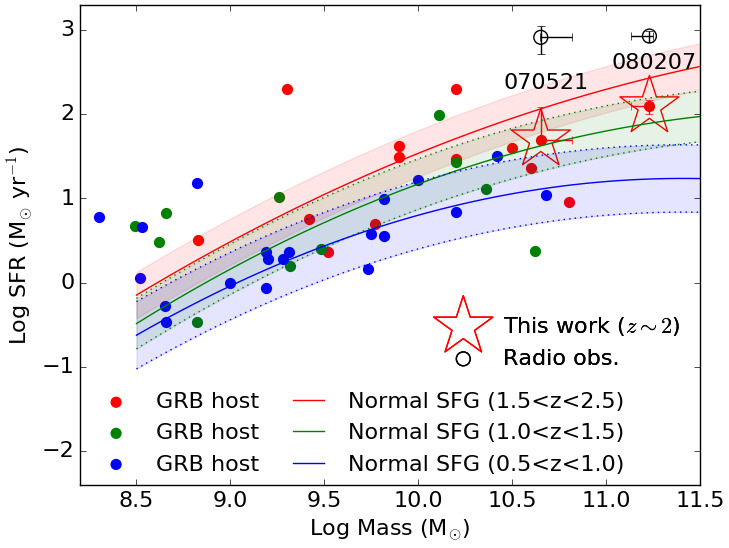}
    \caption{
    Stellar mass as a function of SFR.
    Red dots highlighted by red stars are GRB 070521 and 080207 hosts (this work).
    Other GRB host galaxies are compiled from GHostS project (http://www.grbhosts.org/) and displayed by coloured dots.
    Short GRB host galaxies with $T_{90} < 2.0$ s are excluded, where $T_{90}$ is a time duration of GRB containing 90\% of the total gamma-ray fluence.
    Solid lines are best fit functions of main sequences measured for mass-complete star-forming galaxies from CANDELS fields \citep{Whitaker2014}. %Chabrier2003 IMF%
    For comparison with GRB hosts at $z=2$, main sequences at $1.5<z<2.0$ and $2.0<z<2.5$ are averaged into single redshift bin of $1.5<z<2.5$.
    Shaded regions are $\pm$ 1$\sigma$ dispersion of galaxy distributions.
    The dispersions measured at log $M_{*}$=10.5 are used for each redshift bin \citep{Erfanianfar2016,Mancuso2016}. 
    Colours correspond to different redshift bins.
    SFRs$_{\rm radio}$ of the GRB 070521 and 080207 host galaxies are shown by open circles \citep{Perley2013b}.
    Except for these two, SFR$_{\rm radio}$ is not included in the figure.
    }
    \label{fig4}
\end{figure}

SFRs of star-forming galaxies correlate with various parameters. 
The primary dependent parameter is stellar mass, which is known as the \lq \lq main sequence\rq \rq\ of normal star-forming galaxies. 
Therefore a comparison with the main sequence and the GRB hosts is useful to characterise the GRB host properties.
Fig. \ref{fig4} demonstrates locations of the GRB 070521 and 080207 host galaxies in the main sequence plane. 
We note that the comparison sample displayed by solid lines includes dust-obscured star formation, i.e., $UV+IR$ SFR, by using $Spitzer$/MIPS 24$\mu$m photomery and the conversion to the total IR luminosity \citep{Whitaker2014}. 
However, GRB host sample (coloured dots) compiled from GHostS project\footnote[3]{http://www.grbhosts.org/} is heterogenous in terms of dust obscured SFR.
Many of them lacks SFR$_{\rm IR}$ measurements.
Short GRB host galaxies with $T_{90} < 2.0$ s are excluded from Fig. \ref{fig4}, where $T_{90}$ is a time duration of GRB containing 90\% of the total gamma-ray fluence. 

In Fig. \ref{fig4}, the SFRs of the GRB 070521 and 080207 host galaxies derived from our analysis place them on the main sequence within observational uncertainties.
In terms of molecular gas properties such as SFR/$M_{\rm gas}$, the GRB 080207 host occupies the same location as main sequence galaxies, while the molecular gas excitation is higher than the main sequence \citep{Hatsukade2019}. 
These results support the hypothesis that GRB host properties are more similar to normal star-forming galaxies at the high-$z$ Universe in contrast to low-$z$ \citep[e.g.,][]{Perley2013a}.

SFRs$_{\rm radio}$ of the GRB 070521 and 080207 host galaxies are 871$\pm$300 and 846$\pm$124 $M_{\odot}$ yr$^{-1}$ \citep{Perley2013b}, respectively, which are shown by open circles in Fig. \ref{fig4}.
The SFRs$_{\rm radio}$ are$\sim$7 and 3 times larger than the main sequence at the same redshift range \citep{Whitaker2014}.
These values correspond to $\sim$3 and 2 $\sigma$ above the main sequence, where $\sigma$ is the standard deviation of the galaxy distribution around the main sequence at $1.5<z<2.5$ (red shaded colour in Fig. \ref{fig4}).
A typical dispersion around the SFR$_{\rm FIR}$-SFR$_{\rm radio}$ relation of star-forming galaxies is within a factor of $\sim$1.6 up to $z\sim3$ \citep[e.g.,][]{Bonzini2015}.
Therefore the excess of the SFRs$_{\rm radio}$ from the main sequence is beyond the statistical uncertainty of the SFR estimation.
Our SFRs are significantly lower than that derived from radio observations.
This is probably due to the contamination of radio fluxes from the long-lived afterglows.

There are other outstanding GRB host galaxies from the main sequences. 
These also might be related to the heterogenous SFR measurements.
For instance, high-SFR outliers (e.g., GRB 060814 and 120119A) and low-SFR outlier (061126) lack FIR observations \citep{Perley2008,Morgan2014,Perley2015}.
Because the current sample is too small, we need to wait for future observational data to conclude it.

In summary, the high-SFR$_{\rm radio}$ outliers at $z\sim2$ in the main sequence plane, GRB 070521 and 080207 host galaxies, actually moved on the main sequence, when the SFRs are measured based on the photometries including up to rest-frame FIR.
This result indicates an importance of the rest-frame FIR observations to determine SFR especially for GRB host galaxies to avoid possible effects of long-lived radio afterglows.

\subsection{[C~II]158$\mu$m deficit}
\label{CII}
\subsubsection{$L_{\rm [C~II]}$/$L_{\rm FIR}$ as a function of $L_{\rm FIR}$}
We, for the first time, detected [C~{\sc ii}]\,158\,$\mu$m emission from a GRB host galaxy at $z>2$. 
This is the second detection of [C~{\sc ii}]\,158\,$\mu$m emission among known GRB host galaxies following GRB 980425 \citep{Michalowski2016}.
The [C~{\sc ii}]\,158\,$\mu$m fine structure line is the dominant cooling line of cool interstellar medium, arising from photodissociation regions (PDR) on molecular cloud surfaces.
It is one of the brightest of emission lines from star-forming galaxies from FIR through meter wavelengths, almost unaffected by dust extinction.
[C~{\sc ii}]\,158\,$\mu$m luminosity, $L_{\rm [CII]}$, has been discussed as an indicator of SFR \citep[e.g.,][]{Stacey2010}.
If $L_{\rm [CII]}$ linearly scales SFR, a ratio to FIR luminosity, $L_{\rm [CII]}$/$L_{\rm FIR}$, is expected to be constant, since $L_{\rm FIR}$ is a linear function of SFR \citep[e.g.,][]{Kennicutt1998b}.
However, $L_{\rm CII}$/$L_{\rm FIR}$, is not constant but declines with increasing $L_{\rm FIR}$, known as \lq \lq [C~{\sc ii}] deficit\rq \rq\ \citep[e.g.,][]{luhman1998,Malhotra2001,luhman2003,Sargsyan2012,Diaz2013,Spilker2016,Diaz2017}.
The [C~{\sc ii}] deficit persists when including high-$z$ galaxies \citep[e.g.,][]{Stacey2010,Wang2013,Rawle2014}.
In Fig. \ref{fig5} we compare the [C~{\sc ii}] deficit in the GRB080207 host and other star-forming galaxies.
Two GRB hosts are shown by stars for GRB 080207 (orange star) and 980425 (blue star). 
The comparison sample is compiled from literature up to $z\sim3$ \citep{Cormier2010,Ivison2010,Malhotra2001,Stacey2010,  Sargsyan2012,Cormier2014,Farrah2013,Magdis2014,Brisbin2015,Gullberg2015,Schaerer2015}.
Active galactic nucleus are separated from star-forming galaxies based on either (i) explicit description in the literature or (ii) EW$_{\rm PAH6.2\mu m} < 0.1$ \citep{Sargsyan2012}.
As reported by previous studies \citep[e.g.,][]{Maiolino2009,Stacey2010}, high-$z$ galaxies are located at different place from local galaxies in the $L_{\rm [CII]}$/$L_{\rm FIR}$-$L_{\rm FIR}$ plane.

The GRB 080207 host shows one of the lowest $L_{\rm [CII]}$/$L_{\rm FIR}$ ratios, among star-forming galaxies at $2<z<3$. 
The L$_{\rm [CII]}$/L$_{\rm FIR}$ ratio is as small as that of local star-forming galaxies.
\citet{Stacey2010} found that the [C~{\sc ii}] deficit at $z\sim1-2$ is predominant for AGN-dominated galaxies.
The $L_{\rm [CII]}$/$L_{\rm FIR}$ ratio of AGN-dominated galaxies is $\sim$8 times smaller than that of star formation-dominated galaxies.
Therefore hosting AGN might be a factor to decrease $L_{\rm [CII]}$/$L_{\rm FIR}$. 
So far, no clear evidence of AGN has been reported for GRB 080207 host.
%Actually the observed SED is successfully reproduced by a pure star-forming galaxy template with MIR PAH features which are suppressed if AGN exists (Fig. \ref{fig3}).
As discussed in \citet{Hatsukade2019}, there might be hints of the AGN contribution.
The optical emission line ratios, i.e., BPT diagram \citep{Baldwin1981}, are log([N~{\sc ii}]$\lambda$6584/H$\alpha$)=$-$0.64 and log([O~{\sc iii}]$\lambda$5007/H$\beta$)=0.63 \citep{Kruhler2015}.
The ratios indicate a composite spectrum of AGN and star formation based on a well-defined discrimination for local galaxies \citep{Kewley2006}, while it is still around the boundary of the \lq \lq maximum starburst\rq \rq\ model \citep{Kewley2001}.
The excitation state of molecular gas is higher than that of main sequence galaxies at the same redshift, and is comparable to quasars \citep{Hatsukade2019}.
The extremely high SFR implied from radio flux \citep{Perley2013b} could contain a contamination from an AGN component \citep{Hatsukade2019}.

\subsubsection{$L_{\rm [C~II]}$/$L_{\rm FIR}$ as a function of $\Sigma_{\rm FIR}$}
Apart from AGN contribution, it is suggested that the $L_{\rm [CII]}$/$L_{\rm FIR}$ ratio correlates more tightly with $L_{\rm FIR}$ surface density, $\Sigma_{\rm FIR}$=$L_{\rm FIR}/(\pi r_{\rm FIR}^2)$, than $L_{\rm FIR}$ \citep[e.g.,][]{Diaz2013,Diaz2017}.
\citet{Diaz2013} found that local pure star-forming galaxies selected by 6.2$\mu$m PAH EW $>$ 0.5 $\mu$m show an order of magnitude drop in the $L_{\rm [CII]}$/$L_{\rm FIR}$ ratio as a function of $\Sigma_{\rm FIR}$.
They argued that the decrease of the $L_{\rm [CII]}$/$L_{\rm FIR}$ ratio is a fundamental property of the starburst itself regardless of a rise of AGN activity.
\citet{Spilker2016} demonstrated that dusty star-forming galaxies detected by South Pole Telescope (SPT), high-$z$ star-forming galaxies, and QSOs up to $z\sim 6$ follow the almost same trend as the local GOALS sample \citep{Diaz2013,Diaz2017}, when the [C~{\sc ii}] deficit is plotted as a function of $\Sigma_{\rm FIR}$.
The higher $\Sigma_{\rm FIR}$ implies the stronger heating source of dust, i.e., the stronger UV radiation.
In fact, PDR models predict that a stronger UV radiation field suppresses $L_{\rm [CII]}$/$L_{\rm FIR}$ radio \citep{Stacey2010,Lagache2018}.

These previous works suggest that a primary parameter controlling $L_{\rm [CII]}$/$L_{\rm FIR}$ is $\Sigma_{\rm FIR}$, though AGN could have a minor contribution.
Motivated by this argument, we compare the [C~{\sc ii}] deficit as a function of $\Sigma_{\rm FIR}$ in Fig. \ref{fig6}.
Note that the comparison sample in Fig. \ref{fig6} is not exactly the same as that in Fig. \ref{fig5} due to the availability of $\Sigma_{\rm FIR}$ measurements (see caption for details).
The GRB 080207 host still shows one of the lowest $L_{\rm [CII]}$/$L_{\rm FIR}$ ratios at a fixed $\Sigma_{\rm FIR}$.
The host might be an outlier when compared with galaxies at $z>1.8$ (red dots in Fig. \ref{fig6}), while the the GRB 980425 host is difficult to be compared due to a lack of comparison sample at the same $\Sigma_{\rm FIR}$.

\cite{Smith2017} investigated spatially resolved [C~{\sc ii}] deficits for 54 nearby galaxies.
They demonstrated that the [C~{\sc ii}] deficits depend on the infrared surface brightness. 
The deviations from the $L_{\rm [CII]}$/$L_{\rm FIR}$-$\Sigma_{\rm IR}$ relation are correlated with changes in metallicity.
The higher metallicity shows the lower $L_{\rm [CII]}$/$L_{\rm FIR}$ ratio at a fixed $\Sigma_{\rm IR}$. 
Model calculations of [C~{\sc ii}]\,158\,$\mu$m predict that the deficit is strongly correlated with the strength of radiation field, and that the metallicity is the secondary factor \citep{Lagache2018}.
Therefore the [C~{\sc ii}] deficit found for the GRB 080207 host could be the metallicity effect.

We note that GRB 080207 host could be unusual among known GRB host galaxies in terms of metallicity.
Observationally GRBs tend to happen in low (sub-solar) metallicity environment at least at the low-$z$ Universe \citep[e.g.,][]{Modjaz2008,Levesque2010,Niino2017,Hashimoto2018} as expected from theoretical studies \citep[e.g.,][]{Woosley2006}.
Some GRBs are hosted by metal-rich galaxies \citep[e.g.,][]{Graham2013,Hashimoto2015,Kruhler2015,Stanway2015} in tension with theoretical predictions.
The GRB 080207 host is one of such metal-rich GRB host galaxies.
An optical emission-line diagnostic indicates the metallicity of 12+log(O/H)$_{\rm PP04~N2}$= 8.53$\pm$0.15 based on a calibration by \citet{Pettini2004}.
This is comparable to the Solar value of 8.69 \citep{Asplund2009}.
The upper limit on the [N~{\sc~ii}]\,205\,$\mu$m flux in this work also constrains the metallicity from FIR emission-line diagnostics \citep{Nagao2012}.
The [N~{\sc~ii}]\,205$\mu$m/[C~{\sc~ii}]\,158\,$\mu$m flux ratio provides an upper limit of $\lesssim$1 $Z_{\odot}$, assuming log$n_{\rm e}$=3.0 cm$^{-3}$ and log$U$=$-$2.5, where $n_{\rm e}$ and $U$ are electron density and ionisation parameter, respectively.
The metallicity constrained by FIR fine structure lines is consistent with that from optical emission lines.
\lq \lq Normal\rq \rq\ GRB host galaxies with lower metallicities could show a relatively higher $L_{\rm [CII]}$/$L_{\rm FIR}$ ratio if they also follow the dependence of the [C~{\sc ii}] deficit on metallicity.

Since the metallicity of GRB 080207 host is comparable to that of normal star-forming galaxies at $z\sim 2$ with the same stellar mass \citep[e.g.,][]{Steidel2014}, the [C~{\sc ii}] deficit found for GRB 080207 host might be a new characteristic of GRB host galaxies.
If GRBs originate from  massive stars, they would occur preferentially in star-forming galaxies with top-heavy IMFs.
Actually, the top-heavy IMF is a possible explanation for high [O/Fe] ratios measured for local GRB host galaxies \citep{Hashimoto2018}.
The difference of IMF probably affects on the $L_{\rm [CII]}$/$L_{\rm FIR}$ ratio \citep[e.g.,][]{Lagache2018}.
We note that high-$z$ star-forming galaxies also tend to show top-heavy IMFs \citep[e.g.,][]{Zhang2018}.
We need to wait for future data to conclude the IMF difference between GRB host galaxies and star-forming galaxies at $z\sim1-2$, because the GRB host sample is too small.

\subsubsection{$L_{\rm [C~II]}$/$L_{\rm FIR}$ as a function of $L_{\rm CO(1-0)}$/$L_{\rm FIR}$}
\begin{figure}
	\includegraphics[width=\columnwidth]{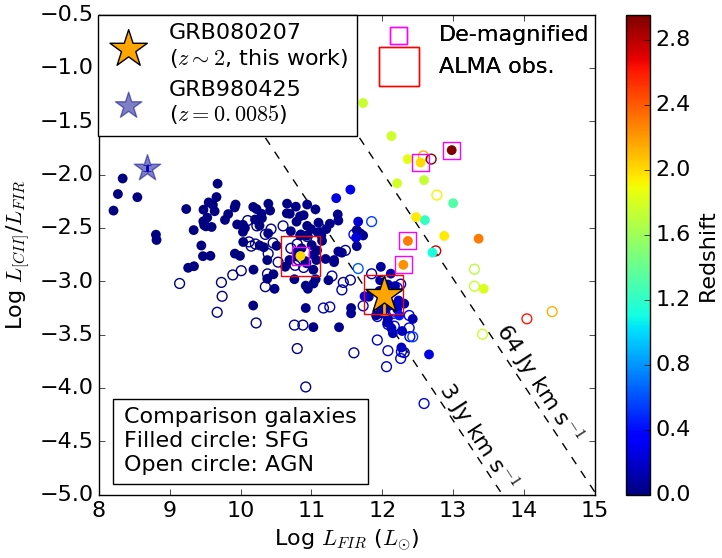}
    \caption{
    $L_{\rm [CII]}$/$L_{\rm FIR}$ as a function of $L_{\rm FIR}$.
    GRB 080207 host (this work) is shown by a orange star, while a blue star indicates a local GRB 980425 host integrated for the entire galaxy \citep{Michalowski2016}.  
    Comparison galaxies at $z<3$ are composed of star-forming galaxies (filled dots) and active galactic nucleus (open dots) compiled from literature \citep{Cormier2010,Ivison2010,Malhotra2001,Stacey2010,  Sargsyan2012,Cormier2014,Farrah2013,Magdis2014,Brisbin2015,Gullberg2015,Schaerer2015}.
    Total IR luminosities in the literature are converted to FIR luminosity by assuming log($L_{\rm TIR}$/$L_{\rm FIR}$)=0.24 \citep{Calzetti2000}.
    Colours from blue to red correspond to redshift, except for squares.
    Two dashed lines are [C~{\sc ii}]\,158$\mu$m detection limits of $f_{\rm [CII]}$=3 and 64 Jy km s$^{-1}$ for galaxies at $z=2$. 
    These values roughly correspond to 3$\sigma$ detection limits of ALMA (Table \ref{tab2}) and ZEUS \citep{Stacey2010}, respectively.
    De-magnified galaxies are marked by magenta squares, which are corrected for magnifications by gravitational lenses.
    ALMA observations are marked by red squares.
    There is no data point behind the legends.
    }
    \label{fig5}
\end{figure}

\begin{figure}
	\includegraphics[width=\columnwidth]{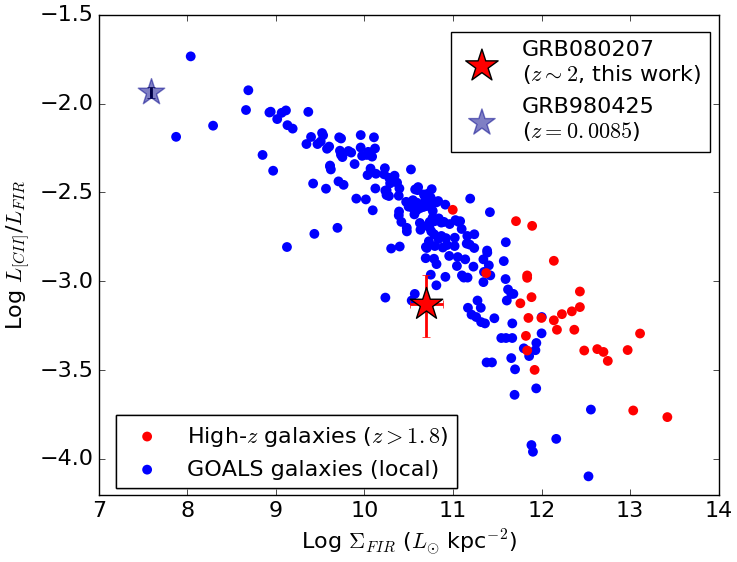}
    \caption{
    $L_{\rm [CII]}$/$L_{\rm FIR}$ as a function of $L_{\rm FIR}$ surface density, $\Sigma_{\rm FIR}$=$L_{\rm FIR}/\pi r_{\rm FIR}^{2}$.
    GRB 080207 host (this work) is shown by a red star. 
    A blue star indicates a local GRB 980425 host galaxy, where we used $L_{\rm FIR}$ of 4.83$\times 10^{8}$ $L_{\odot}$ and size of 2.0 kpc \citep{Michalowski2014,Michalowski2016} to estimate $\Sigma_{\rm FIR}$.  
    Comparison sample is compiled from GOALS project \citep{Diaz2013,Diaz2017} for local galaxies, SPT dusty star-forming galaxies 
    at $z>1.8$ \citep{Spilker2016}, and individual high-$z$ ($z>4$) galaxies and quasars \citep{Walter2009,Carniani2013,Riechers2013,Wang2013,Breuck2014,Neri2014,Riechers2014,Diaz2016,Oteo2016}.
    Total IR luminosities in the literature are converted to FIR luminosity by assuming log($L_{\rm TIR}$/$L_{\rm FIR}$)=0.24 \citep{Calzetti2000}.
    Blue and red colours correspond to local and $z>1.8$ samples, respectively.
    Note that the comparison sample is not exactly the same as that in Fig. \ref{fig5} due to the availability of $\Sigma_{\rm FIR}$ measurements.}
    \label{fig6}
\end{figure}

\begin{figure}
	\includegraphics[width=\columnwidth]{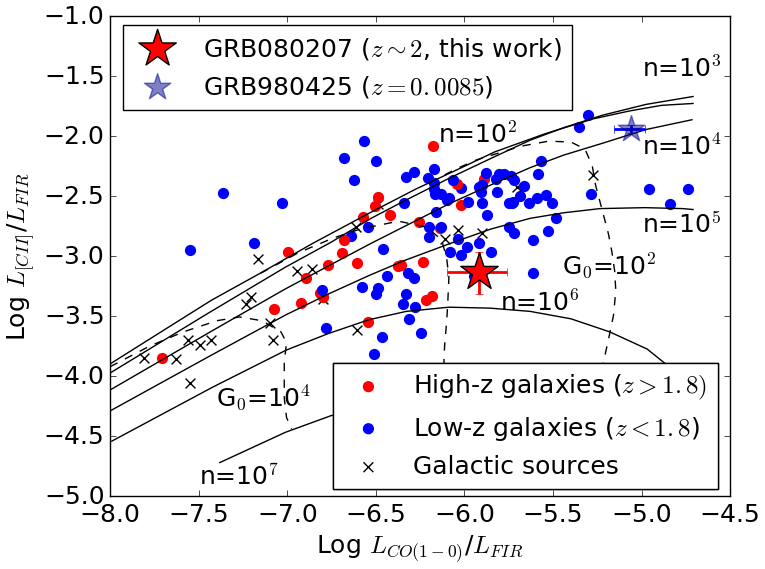}
    \caption{
    $L_{\rm [CII]}$/$L_{\rm FIR}$ as a function of $L_{\rm CO(1-0)}$/$L_{\rm FIR}$.
    GRB 080207 host (this work) is shown by a red star. 
    The CO(1-0) line luminosity of the GRB 080207 host is calculated from the observed CO(1-0) emission-line flux \citep{Hatsukade2019}.
    A blue star indicates a local GRB 980425 host galaxy, where we assumed $L_{\rm CO(1-0)}=2.0\times L_{\rm CO(2-1)}$ \citep{Michalowski2018} to derive $L_{\rm CO(1-0)}$.  
    Comparison sample is compiled from literature including Galactic star-forming regions \citep{Stacey1991}, low-$z$ ($z<1.8)$ galaxies \citep{Stacey1991,Malhotra2001,luhman2003,Cormier2010,Cormier2014,Magdis2014}, and high-$z$ ($z>1.8$) galaxies \citep{Stacey2010,Rawle2014,Schaerer2015,Gullberg2015}.
    Blue and red colours correspond to the low-$z$ and high-$z$ samples, respectively.
    PDR model calculations \citep{Kaufman1999} are shown by solid and dashed lines with different densities, $n$ cm$^{-3}$, and FUV strengths, G$_{0}$.
    Here G$_{0}$ is normalised by $1.6\times 10^{-3}$ erg cm$^{-3}$ s$^{-1}$ \citep{Kaufman1999}.
    Note that the comparison sample is not exactly the same as that in Fig. \ref{fig5} and \ref{fig6} due to the availability of $L_{\rm CO(1-0)}$ measurements. 
    There is no data point behind the legends.
    }
    \label{fig7}
\end{figure}

Gas density is an additional physical parameter to control [C~{\sc ii}] deficit \citep[e.g.,][]{Wolfire1989,Stacey1991,Stacey2010,Hailey-Dunsheath2010,Gullberg2015}.
The critical densities of [C~{\sc ii}]\,158$\mu$m, $n_{\rm crit}$, are $3.0\times 10^{3}$ cm $^{-3}$ for H$^{0}$ collision, and $6.1\times10^{3}$ cm$^{-3}$ for H$_{2}$ collision under a temperature of $T$=100 $K$ \citep{Goldsmith2012}.
Since the [C~{\sc ii}]\,158$\mu$m is a forbidden line, the emission-line luminosity is suppressed in dense environments of $n \gtrsim n_{\rm crit}$.
\citet{Wolfire1989} proposed a diagnostic diagram (Fig. \ref{fig7}) to investigate how FUV-field strength and gas density affect on the [C~{\sc ii}] deficit.
Fig. \ref{fig7} demonstrates $L_{\rm [C~II]}$/$L_{\rm FIR}$ as a function of $L_{\rm CO(1-0)}$/$L_{\rm FIR}$ for the GRB 080207 (red star) and 980425 (blue star) host galaxies along with comparison sample including low-$z$ (blue dots) and high-$z$ (red dots) galaxies.
Note that the comparison sample is not exactly the same as that in Fig. \ref{fig5} and \ref{fig6} due to the availability of $L_{\rm CO(1-0)}$ measurements (see caption for details).
PDR model calculations by \citet{Kaufman1999} are shown by solid and dashed lines, which correspond to different assumptions of gas density, $n$ cm$^{-3}$, and FUV strength, G$_{0}$, respectively.
Here G$_{0}$ is normalised by $1.6\times 10^{-3}$ erg cm$^{-3}$ s$^{-1}$ \citep{Kaufman1999}.
The GRB 080207 host is located between $n$=10$^{5}$ and 10$^{6}$ cm$^{-3}$, corresponding to one of the lowest $L_{\rm [C~II]}$/$L_{\rm FIR}$ ratios at a fixed $L_{\rm CO(1-0)}$/$L_{\rm FIR}$.
Therefore, high gas density is also a possible explanation for the [C~{\sc ii}] deficit of the GRB 080207 host galaxy.
The low strength of the radiation field (G0) seems to explain the high $L_{\rm CO(1-0)}$/$L_{\rm FIR}$ ratios for both the GRB 080207 and 980425 host galaxies.

%[C~{\sc ii}]\,158\,$\mu$m observations of other GRB host galaxies are essentially important to shed light on the new parameter space of GRB host properties.

\subsubsection{Observational bias}
Here we briefly discuss a possible observational bias of ALMA toward a lower $L_{\rm [CII]}$/$L_{\rm FIR}$ ratio.
In Fig. \ref{fig5}, [C~{\sc ii}]\,$\mu$m detection limits of $f_{\rm [CII]}$=3 and 64 Jy km s$^{-1}$ for galaxies at $z=2$ are shown by dashed lines.
These flux limits are adopted from 3$\sigma$ values of [C~{\sc ii}]\,158$\mu$m flux measurements for galaxies at $z\sim2$ observed with ALMA (Table \ref{tab2}) and redshift ($z$) and Early Universe Spectrometer, ZEUS \citep{Stacey2007,Hailey2009,Stacey2010}, respectively.
Therefore the dashed lines roughly correspond to 3$\sigma$ detection limits of ALMA and ZEUS.
Since ALMA observations are deeper than previous [C~{\sc ii}]\,158$\mu$m observations including ZEUS, ALMA can detect the fainter [C~{\sc ii}]\,158$\mu$m emission line while such a population has been missed by previous studies.
Actually both of two galaxies at $z\sim2$ observed with ALMA (large red squares in Fig. \ref{fig5} including this work) indicates lower $L_{\rm [CII]}$/$L_{\rm FIR}$ ratios compared with other galaxies at the same redshift.
One of the two galaxies marked by red square is located beyond the ALMA detection limit, because the observation is practically further deeper due to the gravitational lensing \citep{Schaerer2015}.
Two of other lensed galaxies (magenta) also show relatively lower $L_{\rm [CII]}$/$L_{\rm FIR}$ ratios among $z\sim2$ galaxies.
Therefore the [C~{\sc ii}] deficit of the GRB 080207 host might be the observational bias toward the lower $L_{\rm [CII]}$/$L_{\rm FIR}$ ratio caused by the deep ALMA observation.
Increasing ALMA observations of [C~{\sc ii}]\,158$\mu$m for star-forming galaxies at $z\sim2$ is essentially important to address the possible observational bias of the GRB host galaxy.

\section{Conclusions}
\label{conclusion}
We conducted rest-frame FIR observations of two GRB host galaxies at $z\sim2$ with ALMA band 8 and 9. 
The FIR continuum emissions were detected for the GRB 070521 host with S/N $\sim$8 in band 9, and for the GRB 080207 host with S/N$\sim$19 and 7 in band 8 and 9, respectively.
These detections are complementary with previous photometries by covering the SED peaks of dust emissions heated by star-forming activity, which provide more reliable estimates of dust-obscured SFRs.
The SED fitting analysis indicates SFRs of 49.85 $^{+72.33}_{-2.86}$ and 123.4$^{+25.19}_{-21.78}$ $M_{\odot}$ yr$^{-1}$ for the GRB 070521 and 080207 host galaxies, respectively.
These values are significantly lower than those implied from radio observations ($\sim$800 $M_{\odot}$ yr$^{-1}$).
The derived SFRs are in agreement with the \lq \lq main sequence\rq \rq\ galaxies within the observational uncertainties in the stellar mass-SFR plane, suggesting that these hosts are normal star-forming galaxies at $z\sim2$ rather than starbursts. 
Our results also suggest the possible contamination from the long-lived afterglows even several years after the bursts, which enhanced radio fluxes.

ALMA also detected [C~{\sc ii}]\,158\,$\mu$m emission line from the GRB 080207 host.
This is the first detection of [C~{\sc ii}]\,158\,$\mu$m of a GRB host galaxy at $z>2$, and the second detection among known GRB hosts.
The luminosity ratio of [C~{\sc ii}]\,158\,$\mu$m to FIR, $L_{\rm [CII]}$/$L_{\rm FIR}$, is 7.5$\times 10^{-4}$.
The ratio is one of the lowest values among galaxies at $2<z<3$ with the same $L_{\rm FIR}$, known as \lq \lq [C~{\sc ii}] deficit\rq \rq.
The host could be a [C~{\sc ii}] deficit outlier at $z>1.8$ in the $L_{\rm [CII]}$/$L_{\rm FIR}$-$\Sigma_{\rm FIR}$ plane, where $\Sigma_{\rm FIR}$ is $L_{\rm FIR}$ surface density.
The strong [C~{\sc ii}] deficit found for the GRB 080207 host might be a new physical property to characterise GRB host galaxies at $z\sim1-2$. 
Possible parameters controlling the deficit of the host galaxy include the metallicity, IMF, and gas density. 
In addition, the deep ALMA observations tend to trace the lower $L_{\rm [CII]}$/$L_{\rm FIR}$ ratios.
To fully understand the [C~{\sc ii}] deficit of the GRB host, the possible observational bias needs to be addressed in future observations.
%We note that the metallicity of the GRB 080207 host is one of the highest values of GRB hosts.

\section*{Acknowledgements}
We would like to acknowledge all the staffs at the ALMA Regional Center for their help in preparation for ALMA observations and data acquisition.
We are very grateful to the anonymous referee for many insightful comments.
TG acknowledges the supports by the Ministry of Science and Technology of Taiwan through grants 105-2112-M-007-003-MY3 and 108-2628-M-007-004-MY3.
BH is supported by JSPS KAKENHI grant No. 19K03925.
This research has made use of the GHostS database (www.grbhosts.org), which is partly funded by Spitzer/NASA grant RSA Agreement No. 1287913.
This paper makes use of the following ALMA data: ADS/JAO.ALMA\#2017.1.00337.S. 
ALMA is a partnership of ESO (representing its member states), NSF (USA) and NINS (Japan), together with NRC (Canada), MOST and ASIAA (Taiwan), and KASI (Republic of Korea), in cooperation with the Republic of Chile. 
The Joint ALMA Observatory is operated by ESO, AUI/NRAO and NAOJ.
This research has made use of the SVO Filter Profile Service (http://svo2.cab.inta-csic.es/theory/fps/) supported from the Spanish MINECO through grant AYA2017-84089.

%%%%%%%%%%%%%%%%%%%%%%%%%%%%%%%%%%%%%%%%%%%%%%%%%%



\bibliographystyle{mnras}
\bibliography{GRB_ALMA_MNRAS} % if your bibtex file is called example.bib

%\begin{thebibliography}{99}
%\end{thebibliography}

%%%%%%%%%%%%%%%%%%%%%%%%%%%%%%%%%%%%%%%%%%%%%%%%%%

%%%%%%%%%%%%%%%%% APPENDICES %%%%%%%%%%%%%%%%%%%%%
%\appendix
%\section{Images of member galaxies in the blue cluster}

% Don't change these lines
\bsp	% typesetting comment
\label{lastpage}
\end{document}